\documentclass[12pt,a4paper]{article}

\usepackage{amsmath,amssymb,amsthm,amsfonts}
\usepackage[dvips]{graphicx}
\usepackage{epsfig}
\usepackage{siunitx}	
\usepackage{subfigure}
\usepackage{float}
\usepackage{verbatim}
\usepackage[font=small]{caption}
\usepackage{soul}

\usepackage{color}

\title{Mathematical modelling of fibre coating}
\author{\textbf{Group members:}\\ Maria Aguareles\footnote{Department d'Informatica, Matematica Aplicada i Estadistica, Universitat de Girona, Girona, Spain}, Francesc Font\footnote{Centre de Recerca Matem\`{a}tica, Campus de Bellaterra  Edifici C, 08193 Bellaterra, Barcelona, Spain.}, Tim Myers\footnotemark[2], Jordi Ripoll\footnotemark[1]}

\def\p{\partial}

\def\({\text{\huge (}}
\def\){\text{\huge )}}
\def\bf{\textbf}
\def\]{\text{\huge ]}}
\def\[{\text{\huge [}}

\def\bf{\textbf}

\newcommand{\bi}{\begin{itemize}}
\newcommand{\ei}{\end{itemize}}
\newcommand{\be}{\begin{equation}}
\newcommand{\ee}{\end{equation}}
\newcommand{\ba}{\begin{align}}
\newcommand{\ea}{\end{align}}

\newcommand\nc{\newcommand}
\nc\pad[2]{\frac{\p #1}{\p #2}} \nc\padd[2]{\frac{\p^2 #1}{\p
{#2}^2}} \nc\nd[2]{\frac{d #1}{d #2}} \nc\pat[2]{\frac{D #1}{D
#2}} \nc\ov{\overline} \nc\degree{^{\circ}} \nc\ord[1]{{\cal
O}(#1)} \nc\ra{\rightarrow} \nc\Ra{\Rightarrow} \nc\dint{{\mbox ~
d}}

\newcommand{\bea}{\begin{eqnarray}}
\newcommand{\eea}{\end{eqnarray}}
\newcommand{\beas}{\begin{eqnarray*}}
\newcommand{\eeas}{\end{eqnarray*}}

\begin{document}
\maketitle

\begin{abstract}
In this report we formulate and analyse a mathematical model describing the evolution of a thin liquid film coating a wire via an extrusion process. We consider the Navier-Stokes equations for a 2D incompressible Newtonian fluid coupled to the standard equation relating the fluid surface tension with the curvature. Taking the lubrication theory approximation and assuming steady state, the problem is reduced to a single third-order differential equation for the thin film height. An approximate analytical solution for the final film height is derived and compared with a numerical solution obtained by means of a shooting scheme. Good agreement between the two solutions is obtained, resulting in a relative error of around 5\%. The approximate solution reveals that the key control parameters for the process are the initial film height, the fluid surface tension and viscosity, the wire velocity and the angle of exit at the extruder.
\end{abstract}

\section{Introduction}

It is anticipated that in the near future toxic components, such as toluene and sepiolite, will be prohibited in the manufacturing process used by Frenos Sauleda S.A to fabricate clutch components. Toluene is already listed as a restricted substance by the European Chemicals Agency \cite{ECA} and sepiolite, although not restricted by ECA yet, can be potentially carcinogenic depending on the characteristic size of its fibres \cite{Bellmann}. The study group was asked to investigate the process and come up with recommendations to ensure the quality of the final product after the removal of these components.

\begin{figure}[ht]
  \centering
  \includegraphics[width=12cm]{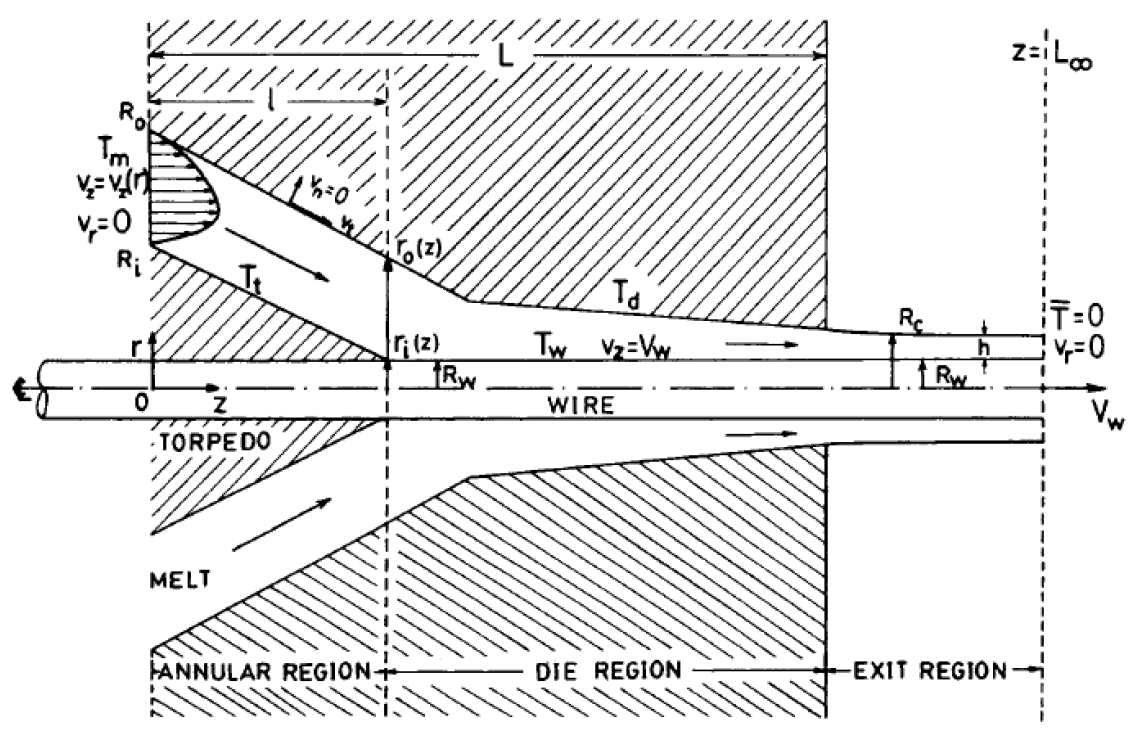}
  \caption{Schematic of wire coating process (reproduced from \cite{Mitsoulis} with permission from Wiley). The current report focusses on the prediction of the film height $h$ in the exit region.
  }
  \label{fig_schematics}
\end{figure}

A key stage in the process is the coating of wires via extrusion. Three wires enter the extruder, these are then entangled and coated to form a single component. Upon leaving the extruder the composite wire should have a specified coating thickness and/or weight. An illustration of a typical wire coating via extrusion process is presented in Figure~\ref{fig_schematics}. During the meeting, work focussed on this wire coating process.

The coating material starts as a wet powder which is fed into a hopper and subsequently heated so that certain components melt, resulting in a thick paste similar to plasticine. The paste is composed of a number of components, including plastic. This together with the sepiolite suggests that the paste will have significant non-Newtonian behaviour. A Newtonian fluid exhibits a constant viscosity when subjected to a shear stress, non-Newtonian fluids can exhibit many unusual behaviours such as shear thinning or thickening, elasticity or a yield stress (below this they behave as a solid, above as a liquid).

The role of the toluene, a volatile solvent, is to reduce the viscosity during the coating process but then rapidly evaporate over a very short distance (of the order of centimetres) after leaving the extruder. Sepiolite is a fibrous clay mineral with good adsorbent properties. The fibrous structure can add strength to a solid material, in a liquid it can result in thixotropic (time-dependent shear thinning) behaviour.

The process of coating a solid surface with a fluid is relevant in many industrial applications. The process typically involves the withdrawal of a solid from a liquid bath, and is sometimes referred to as \textit{the drag-out problem}. The key aspect in the drag-out problem is to predict the final height of the coating film and find out the most relevant physical parameters involved in the process. There exists a number of interesting reviews on the specific topic of coating \cite{Que,Wei} and, in more general terms, on thin liquid films \cite{Cras,Myers98}. Nowadays, there are whole scientific journals dedicated to coating research \cite{Coa,Sur}. Probably the first theoretical studies describing the coating process were carried out in the 1940s by Derjaguin \cite{Der} and Landau \& Levich \cite{Lan}, in an attempt to mathematically model the deposition of a photosensitive emulsion on a solid surface for cinefilm manufacture \cite{Lan}. An extension of the theory was introduced by White and Tallmadge \cite{Whi} which allowed to predict the evolution of the film thickness for a wider variety of coating fluids. Later, the drag out problem was analysed in more detail and solutions to the governing equations were obtained by means of matched asymptotic expansions \cite{Jam,Wil,Tuck}.

In this report, in order to predict the final thickness of the coating used in Frenos Sauleda wires we will take an approach similar to that in Landau \& Levich \cite{Lan}. The outline of the report is as follows. In Section \ref{mod} we describe the main assumptions we make to describe the coating process and formulate the mathematical model. In Section \ref{num}, we discuss the strategy to solve the model numerically. In Section \ref{SecAprox}, we provide an approximate asymptotic solution for the final thickness of the coating film and compare it with the numerical prediction. In Section \ref{results} we present some numerical simulations and compare these results with the approximate solution obtained in Section \ref{SecAprox}. Finally, in Section 5 we present our conclusions and provide a list of possible future routes to improve the present wire coating manufacturing process.

\section{Mathematical model}
\label{mod}
When the wire leaves the extruder it drags-out the coating fluid as illustrated in Figure~\ref{fig_schematics}. The equations governing the fluid dynamics are the Navier-Stokes equations which are to be coupled to surface tension experienced at the fluid free surface ($y=h(x,t)$) and to the wire velocity at $y=0$ through the boundary conditions. Some material properties and relevant parameter values for the current industrial process are listed in Table \ref{Table:PhysicalParameters}.

Solving the Navier-Stokes equations in their most general form is very challenging. In order to reduce the complexity of the problem we make the following assumptions:
\begin{itemize}
    \item[a)] We assume a flat surface (it is shown in Appendix B that this leads to errors of the order 1\%).
    \item[b)] We consider the coating fluid to behave as a Newtonian incompressible fluid.
    \item[c)] We neglect gravity effects.
    \item[d)] Since the ratio of the initial film thickness to the coated fibre radius is small (approximately 0.07), the coating layer can be considered a thin film.
\end{itemize}
After these assumptions are made the Navier-Stokes equations are substantially reduced (see equations \eqref{pde1}-\eqref{pde2} in Appendix A) and an expression for the fluid velocity can be obtained (see equation \eqref{u} in Appendix A).

On a flat surface the drag-out problem is governed by a standard mass balance:
\begin{equation}\label{eq_h}
\frac{\partial h}{\partial t} + \frac{\partial Q}{\partial x} = 0\,,
\end{equation}
where $h(x,t)$ is the height of the liquid layer and $Q(x,t)$ is the fluid flux per unit length. A derivation of equation \eqref{eq_h} can be found in Appendix A.
Following the derivation and the notation in the Appendix we find the flux  is given by
\bea
Q = \int_0^h u ~ dz = \frac{\sigma}{3\mu} h^3 h_{xxx} + U h ~ ,
\eea
where $\sigma$ is the surface tension, $\mu$ the viscosity, $U$ the speed of the fibre, $u$ the fluid velocity, and the subscript $x$ indicates the derivative of $h(x,t)$ with respect to $x$. Hence
\begin{equation}\label{eq_h2}
\frac{\partial h}{\partial t} + \frac{\partial }{\partial x}\left(\frac{\sigma }{3\mu} h_{xxx}h^3+Uh\right) = 0 ~ .
\end{equation}
This equation shows that the height of the fluid layer varies due to the effect of surface tension (more precisely the ratio of surface tension to viscosity) and the pulling speed.

\begin{figure}
\begin{center}
\includegraphics[scale=0.35]{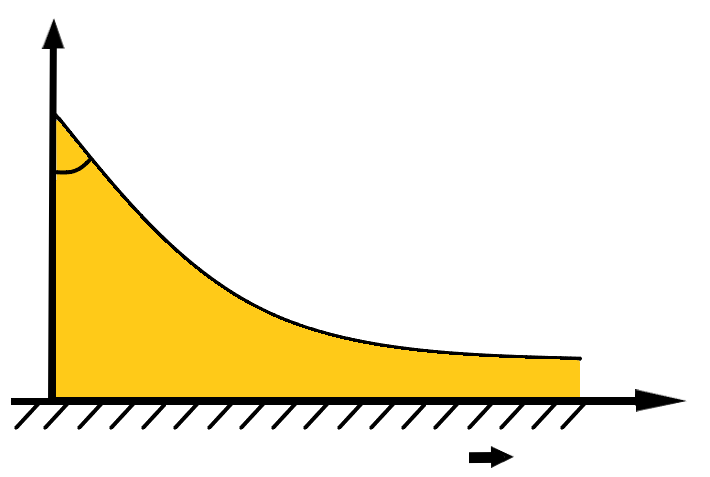}
\put(-190,95){$h_0$}
\put(-135,60){$h(x)$}
\put(-10,10){$x$}
\put(-25,30){$h_{\infty}$}
\put(-185,120){$y$}
\put(-168,70){$\theta$}
\put(-48,0){$U$}
\put(-100,0){Fibre}
\put(-160,30){Liquid}
\end{center}
\caption{\label{fig:angle} Sketch of the fluid exiting the extruder.}
\end{figure}
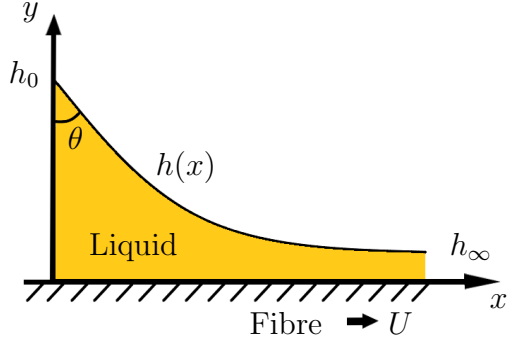

Most drag-out processes are continuous, running at a constant speed. Mathematically this means we may treat the process as steady-state, hence we neglect the time derivative in both \eqref{eq_h} and \eqref{eq_h2} to find an ordinary differential equation for the flux. If we note that far from the outlet the height is constant and denote it by $h_\infty$ then the far-field flux is $Q=U h_{\infty}$ and integrating the steady form (i. e. $\partial h/\partial t=0$) of \eqref{eq_h2} leads to
\begin{equation}
\label{hxxx}
h_{xxx} = \frac{3 \mu U}{\sigma}\frac{(h_{\infty}-h)}{h^3}\,.
\end{equation}
This is a standard form of equation, applicable to a multitude of physical processes. Various forms, both steady and unsteady, are discussed in \cite{Myers98}. The (non-dimensional) ratio $\mu U/\sigma$ is known as the capillary number and represents the relative importance of viscous drag to surface tension forces.
Since equation \eqref{hxxx} is third-order we expect three boundary conditions, however it is autonomous (invariant in shifts in $x$) which reduces the required conditions to two. Below we write down three but the $h_{\infty}$ condition has already been imposed
\bea
\label{hBCs}
\lim_{x \rightarrow \infty} h = h_{\infty}, \qquad h(0)=h_0, \qquad \nd{}{x} h(0) = -  \cot \theta\,,
\eea
where $h_0$ and $\theta$ are the thickness and exit angle of the fluid as it leaves the extruder as shown in Figure \ref{fig:angle}.

\begin{table}[ht]
\centering
\begin{tabular}{ c  c  c  c }
\hline
Quantity & Symbol & Value & Units \\
\hline      
Liquid density   & $\rho$   & 2000              & \SI{}{kg.m^{-3}}\\
Liquid viscosity & $\mu$    &  10               & \SI{}{Pa . s} \\
Surface tension  & $\sigma$ &  30$\cdot10^{-3}$ & \SI{}{N.m^{-1}} \\
Gravity          & $g$      & 9.81              & \SI{}{m.s^{-2}} \\
Fibre velocity   & $U$      & 2                 & \SI{}{m.s^{-1}}\\
Fibre radius     & $R$      & 0.003                 & \SI{}{m}\\
Initial film thinckness    & $h_0$      & 0.0002                 & \SI{}{m}\\
\hline
\end{tabular}
\caption{Typical parameter values required for the model. }
\label{Table:PhysicalParameters}
\end{table}

\subsection{Dimensionless model}

Mathematically it is simplest to work in terms of a dimensionless model. We denote $\hat{x} = x/L$ and $\hat{h}(\hat{x}) =h(L\hat{x})/ h_{\infty} $. The height-scale $h_{\infty}$ is chosen as the final thickness, the length-scale is as yet unknown. The governing equation \eqref{hxxx} becomes
\begin{equation*}
\frac{h_{\infty}}{L^3}
\hat{h}_{\hat{x}\hat{x}\hat{x}} = \frac{3 \mu U}{\sigma h_{\infty}^2}
\frac{(1-\hat{h})}{ \hat{h}^3} ~,
\end{equation*}
which indicates the length-scale $L = h_{\infty} \sqrt[3]{\sigma/(3\mu U)}$. The governing equation and boundary conditions may now be written
\begin{equation}
\label{h}
\hat{h}_{\hat{x}\hat{x}\hat{x}} = \frac{1-\hat{h}}{ \hat{h}^3} ~,
\end{equation}
which we solve subject to
\bea
\lim_{\hat{x} \rightarrow \infty} \hat{h}(\hat{x}) = 1 \qquad \hat{h}(0)=\hat{h}_0=\frac{h_0}{h_{\infty}} \qquad \hat{h}_{\hat{x}}(0) = - \frac{L}{h_{\infty}} \cot \theta=-\cot{\theta}\sqrt{\frac{\sigma}{3\mu U}}~.
\eea

An important feature of the non-dimensional formulation is that it clearly shows the dependence of the problem on the physical conditions. Now we can see that there are only two important parameter values $\hat{h}_0= h_0/h_{\infty}$ and $\sqrt[3]{\sigma/(3\mu U)} \cot \theta$. The physical meaning of this is that we may operate at different conditions but achieve the same final film height if we keep these parameters constant. This is key to the present study, if we remove toluene we increase viscosity: this formula indicates how to compensate for this. For example, if all other conditions are kept the same, then increasing $h_0$ will lead to the same increase in $h_{\infty}$, or say increasing the viscosity by a factor 2 we may achieve the same final film height by decreasing the speed by a factor 2.

\section{Numerical solution}
\label{num}
The numerical problem is easily solved using a marching scheme and shooting. The starting point is based on the fact that sufficiently far from the outlet the height is approximately $h_{\infty}$, that is
\bea
\lim_{\hat{x} \rightarrow \infty} \hat{h}(\hat{x}) = 1
\eea
so we may write $\hat{h}(\hat{x}) = 1 + \epsilon f(\hat{x})$ where $\epsilon \ll 1$ and $f(\hat{x})$ is an unknown function. Substituting into the governing equation \eqref{h}
\bea
\epsilon f_{\hat{x}\hat{x}\hat{x}} = \frac{1-(1+\epsilon f)}{(1+\epsilon f)^3} = -\epsilon f (1-3 \epsilon f + \cdots) ~ .
\eea
We can choose $\epsilon$ arbitrarily small, provided we move sufficiently far from the outlet, which also corresponds to linearising equation \eqref{h} about the constant solution $\hat{h}\equiv 1$. Hence we may neglect terms of order $\epsilon$ and find
\bea
f_{\hat{x}\hat{x}\hat{x}} = - f ~ .
\eea
This equation has three solutions of the form $f = e^{m\hat{x}}$, where $m^3=-1$. There is one real solution $m=-1$ and two complex ones. The complex solutions lead to an oscillating film. Physically this occurs when the coated wire enters into the outlet rather than leaving it (i.e. the velocity $U$ is negative). In the present case we deal with the drag out problem, so the film decreases smoothly in height, hence only the solution with $m=-1$ is physical.
Consequently, far from the outlet the height varies according to
\bea
\hat{h}(\hat{x};\epsilon) = 1 + \epsilon e^{-\hat{x}} ~.
\eea
This is the form we use to start the marching scheme.
We therefore solve equation \eqref{h} in a bounded domain $\hat{x}\in[0,\ell]$, with initial conditions
\bea
\label{ICNum}
\hat{h}(\ell;\epsilon)=1+\epsilon e^{-\ell},\quad \hat{h}_x(\ell;\epsilon)=-\epsilon e^{-\ell},\quad \tilde{h}_{xx}(\ell;\epsilon)=\epsilon e^{-\ell}~.
\eea
In practice we fix $\ell \gg 1$ and choose $\epsilon$ so that $\epsilon e^{-\ell}$ is small. We then use $x=\ell$ as the initial point from which we march backwards until $x=0$. For a fixed $\ell$, the idea is therefore to find a specific $\epsilon= \epsilon^*$ such that
\bea
\label{dh0}
\hat{h}_{\hat{x}}(0;\epsilon^*)=-\cot{\theta}\sqrt{\frac{\sigma}{3\mu U}}~.
\eea
We note that the gradient boundary condition depends only on physical parameters of the problem so \eqref{dh0} provides an equation for $\epsilon$ which we solve using the Newton-Raphson method. Finally, to obtain $h_\infty$ we use that $h_0$ is the initial film thickness, which is a known physical quantity, and therefore
\bea
h_\infty=h_0/\hat{h}_0~,
\eea
where $\hat{h}_0=\hat{h}(0;\epsilon^*)$. In Section \ref{results} we present some results obtained from these numerical simulations.

\section{Approximate solution}
\label{SecAprox}
Returning to the approximate solution employed to start the marching scheme we assume that there is only a small change in height from the outlet to the far-field. In which case the approximate solution may be applied throughout the domain. Equating the approximate solution with the angle at $\hat{x}=0$ then determines $\epsilon = L \cot \theta/h_{\infty}$ and so
\bea
\hat{h}(\hat{x}) = 1 + \frac{L}{h_{\infty}} \cot\theta ~ e^{-\hat{x}} ~ .
\eea
At $\hat{x}=0$ we find
\bea
\hat{h}_0 = 1 + \sqrt[3]{\frac{\sigma}{3 \mu U}} \cot \theta  ~ .
\eea
Returning to dimensional form and rearranging gives an expression for the final film height
\bea
\label{aprox}
h_{\infty} = \frac{h_0}{1-\sqrt[3]{\sigma/(3 \mu U)}\cot \theta } ~ .
\eea
In the following section we shall show the good agreement between this expression and the values obtained through direct numerical simulations.

\section{Results}
\label{results}
\begin{figure}[ht]
    \centering
   \fbox{\includegraphics[scale=0.5]{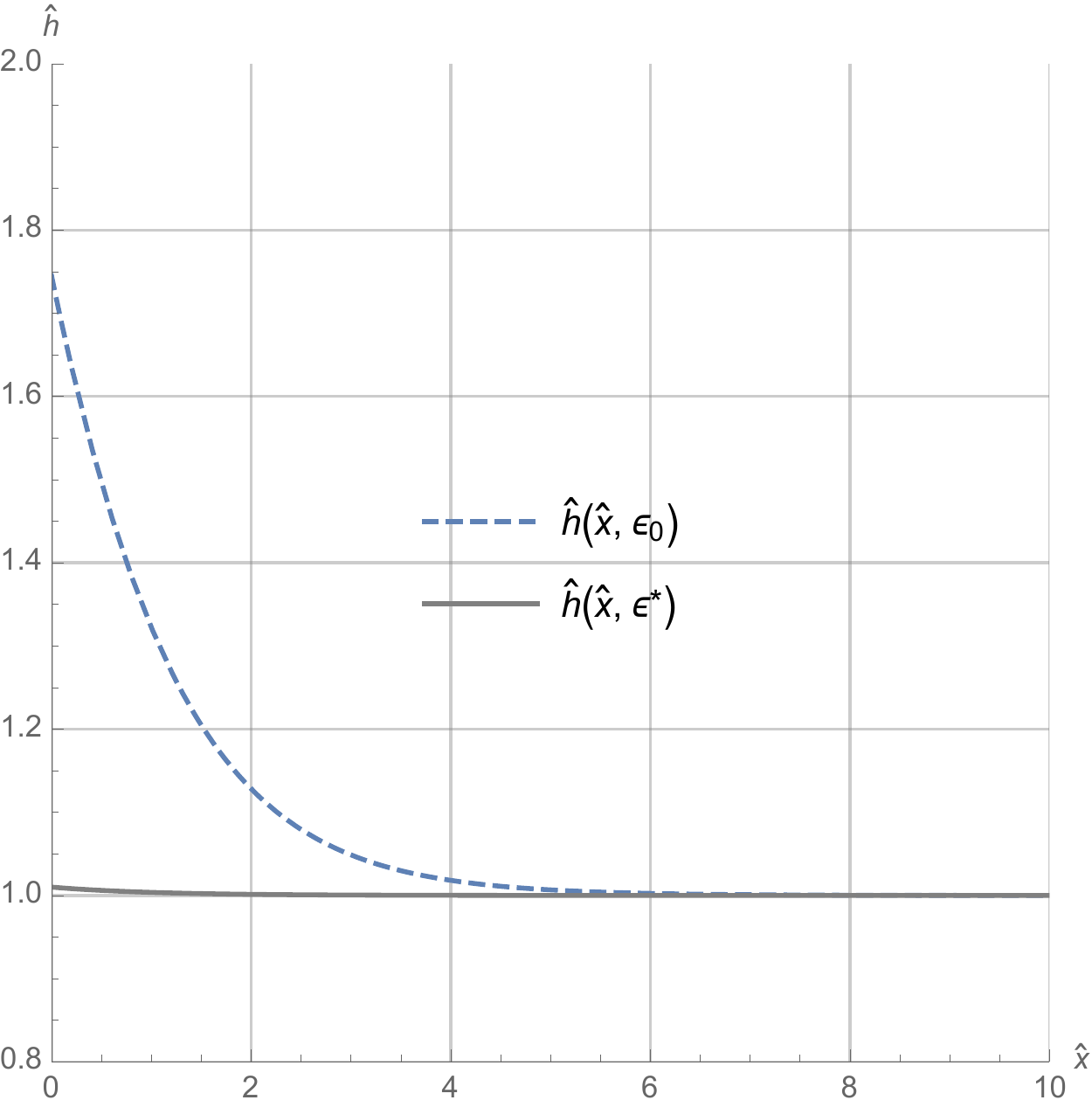}}
   \fbox{\includegraphics[scale=0.5]{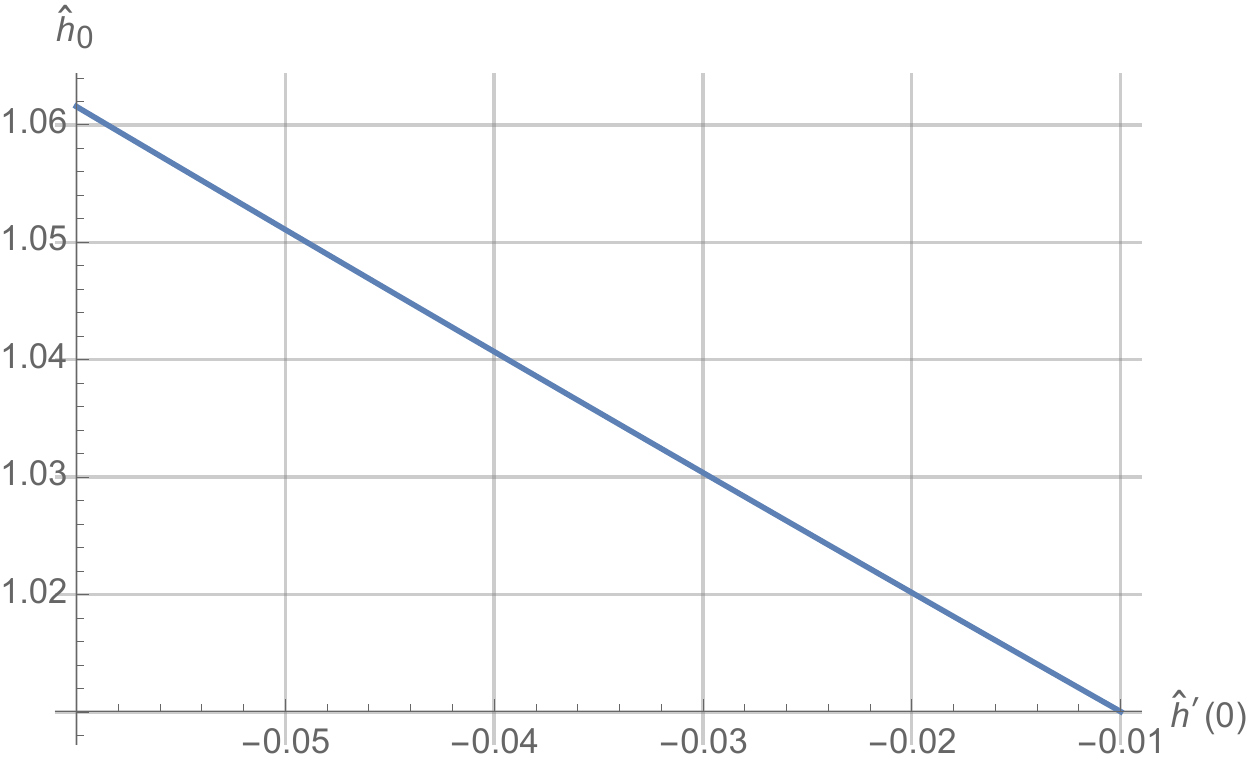}}
    \caption{Left: $\hat{h}_0(\hat{x};\epsilon_0)$ and $\hat{h}_0(\hat{x};\epsilon
^*)$ for $\ell=10$, $\epsilon_0=1$, $\epsilon^*=0.009774$ for $\hat{h}'_0(0;\epsilon^*)=-0.01$. Right: }
    \label{fig:shooting}
\end{figure}

As explained in Section \ref{num}, we start our shooting scheme by fixing $\ell$ and an initial value for the shooting parameter, $\epsilon=\epsilon_0$, which, by marching backwards from $x=\ell$ with \eqref{ICNum} as initial conditions, provides an initial profile for $\hat{h}_0(\hat{x};\epsilon_0)$. Newton-Raphson's method then provides the value of $\epsilon^*$ such that $\hat{h}_{\hat{x}}(0;\epsilon^*)$ achieves the prescribed value given in \eqref{dh0}.  As a way of illustration, in Figure \ref{fig:shooting}-left we fix $\ell=10$ and show in the same graph the initial profile, $\hat{h}_0(\hat{x};\epsilon_0)$ and the final one $\hat{h}_0(\hat{x};\epsilon^*
)$, where the gradient boundary conditions at $\hat{x}=0$ is satisfied.

For each choice of $\hat{h}_{\hat{x}}(0)$ a different initial height $\hat{h}_0$ is obtained. To choose realistic values for $\hat{h}_{\hat{x}}(0)$ we consider the typical parameter values shown in Table \ref{Table:PhysicalParameters} and substitute in equation \eqref{dh0} considering angles $\theta$ ranging from $\pi/8$ to $3\pi/8$. The results are shown in Figure \ref{fig:shooting}-right.

\begin{figure}[ht]
    \centering
    \fbox{\includegraphics[scale=0.5]{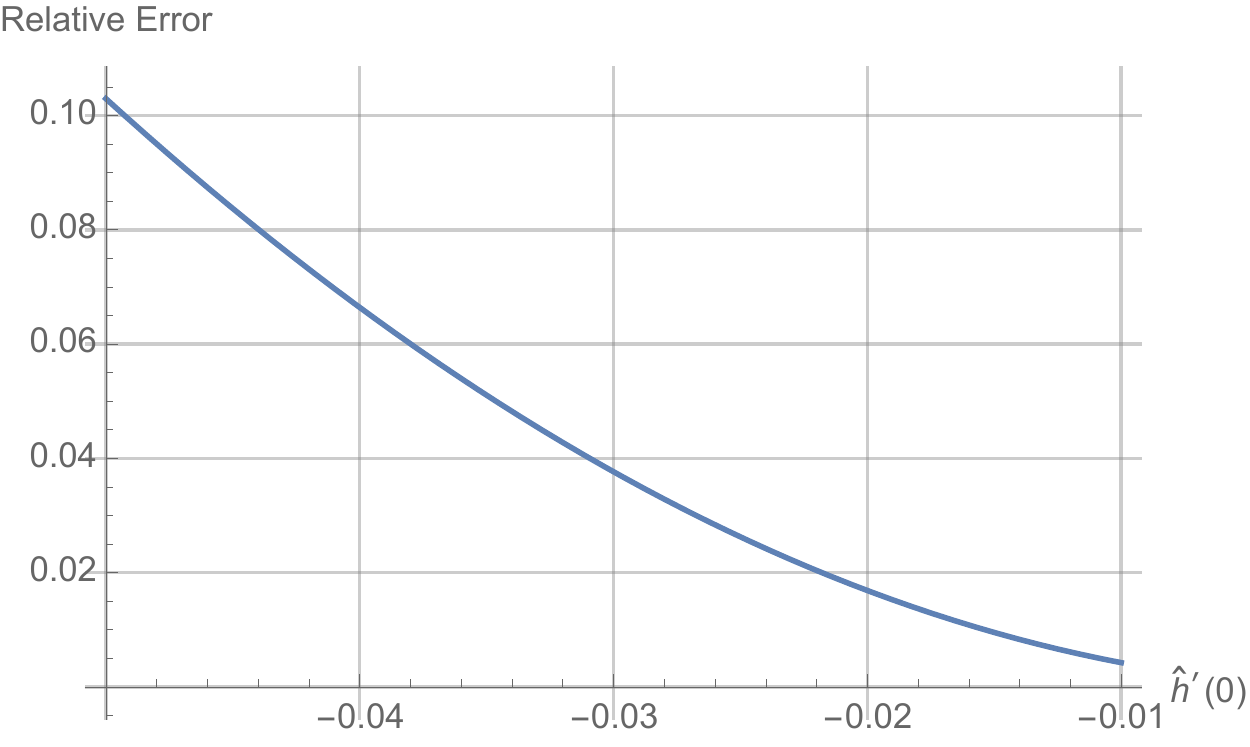}}
    \caption{Relative error (in percentage form) for the values of $h_\infty$ obtained with the approximate expression \eqref{aprox} with respect to the numerically obtained one for different initial gradient values $\hat{h}_{0\hat{x}}(0)$.}
    \label{fig:comp}
\end{figure}

In terms of applicability in an industrial setting, it is obvious that the approximate expression \eqref{aprox} would be more useful than having to go through numerical simulations to compute $h_\infty$. In Figure \ref{fig:comp} we show the good agreement between the approximate formula and the exact values numerically obtained, which justifies the use of the latter. We observe that for values of the initial gradient $\hat{h}_{0\hat{x}}(0)$ between -0.05 and -0.01, which correspond to exit angles from $3\pi/8$ to $\pi/8$, the maximum relative error is of 10\% and it takes place for the angle of $3\pi/8$, which, according to the sketch in figure \ref{fig:angle}, it would correspond to a fluid exiting the extruder almost horizontally.

\section{Conclusion}
\label{conc}

In this report we have developed a basic model to determine the variation of the thickness of the paste as it emerges from the extruder and is carried along with the wire. It is a simple matter to solve the equation numerically, however the approximate analytical solution shows that the final paste height is well approximated by
\begin{equation*}
    h_{\infty} \approx \frac{h_0}{1+\sqrt[3]{\frac{\sigma}{3 \mu U}} \, \cot \theta} ~.
 \end{equation*}
From this we can see the controlling parameters for the process: the initial height, surface tension, viscosity, velocity and angle of exit at the extruder. The final component, $\cot \theta$, depends on the internal dynamics of the extruder and so are beyond this study. The other parameters are simpler to control. The final product weight may be estimated by assuming the three wires used in the process may be represented by an equivalent wire of radius of $R$, then
$$ W \approx \rho_p g \pi \left[ (R+h_{\infty})^2 - 3 R_w^2 \right] + 3 \rho_w g \pi  R_w^2\, ~,~ \qquad R \approx 2 R_{w} ~ ,
$$
where $R_w$ is the actual wire radius.

The analytical solutions permit various observations, specifically we can see how to retain a layer of the same height/weight by adjusting different parameters. For example,
if we replace toluene resulting in a  decrease in $\mu$ but keep $\sigma$ fixed we can still achieve the same final thickness by increasing $U$  accordingly. If the toluene is all evaporated its replacement by another volatile solvent won't affect the final weight. On the other hand
sepiolite remains in final product so a replacement should be of a similar density.

Future work:
\\
For the academics
\begin{itemize}

\item Extend the model to include variable fluid properties (for example the paste is non-Newtonian);
\item Include evaporation  (preliminary calculations suggest a 13\% decrease in height?);
\item Investigate the  existing wide literature on wire extrusion, including non-Newtonian effects.
\end{itemize}
For Frenos Sauleda
\begin{itemize}
\item Carry out detailed experiments to determine the properties of the melted paste and also to measure angle $\theta$, which is crucial to the results and depends on paste and flow conditions.
\end{itemize}

\section*{Appendix A: General equation for the film height}

For the steady flow of a thin layer of fluid being dragged along on a fibre the dominant terms in the Navier-Stokes equations are
\begin{align}
0 &= -\frac{\partial p}{\partial x} + \mu \frac{\partial^2 u}{\partial y^2} \label{pde1}\\
0 &=-\frac{\partial p}{\partial y}   \label{pde2} ~ ,
\end{align}
where $p(x,y)$ is the pressure and $u(x,y)$ is the horizontal component of the velocity at each point $(x,y)$ of the fluid. These require solving subject to the following boundary conditions
\begin{align}
u = U\,,\quad v = 0\quad &\text{at}\ y = 0 \label{bc1}\\
\frac{\partial u}{\partial y} = 0\,, \quad p = -\sigma \kappa \quad &\text{at}\ y = h(x,t)  ~ .\label{bc2}
\end{align}
On the surface $y=0$ the fluid in contact with the fibre matches the fibre velocity $U$, with no vertical velocity. At the free surface $y=h(x,t)$ there is no sheer stress and the vertical stress is due to surface tension, where $\sigma$ is the surface tension and the curvature
\begin{equation}
\kappa = \frac{h_{xx}}{\left(1+h_x^{2}\right)^{3/2}}
\end{equation}
where the subscript in $h_{x}$ indicates derivative with respect to $x$. Equations (\ref{pde1}, \ref{pde2}) are consistent with lubrication theory, which is based on the the assumption the height is significantly less than the length, in line with this we find $\kappa \approx h_{xx}$.

Equation \eqref{pde1} indicates $p=p(x)$ is a function of $x$ alone. The boundary condition then indicates $p = \sigma \kappa$ everywhere.
Equation \eqref{pde2} then becomes
\begin{equation}\label{pde3}
\sigma h_{xxx} + \mu \frac{\partial^2 u}{\partial y^2} = 0\, .
\end{equation}
Integrating \eqref{pde3} twice and applying the boundary conditions \eqref{bc1}-\eqref{bc2} we obtain
\begin{equation}\label{u}
u = \frac{\sigma}{2\mu} h_{xxx}\, y (2h-y) + U\,.
\end{equation}

The kinematic condition
\begin{equation}\label{kine}
\frac{d}{dt}(y(t)-h(x(t),t)) = 0 \implies v(h) =  u(h) \frac{\partial h}{\partial t} + h_x\, ,
\end{equation}
is the mathematical statement that fluid particles on the surface stay there.
For an incompressible fluid
\begin{equation}\label{incom2}
\frac{\partial u}{\partial x} + \frac{\partial v}{\partial y} = 0\,.
\end{equation}
We may integrate this between $y=[0,h]$
\begin{equation}\label{}
v(h) = -\int_{0}^{h}\frac{\partial u}{\partial x}\,dy\,
\end{equation}
after noting $v(0)=0$. Combining this with the kinematic condition
\begin{equation}\label{comb}
u(h)h_x+ \frac{\partial  h}{\partial  t} = -\int_{0}^{h(x)} \frac{\partial u}{\partial x} \,dy ~.
\end{equation}
Leibniz theorem shows that
\bea
\pad{}{x} \int_0^h u ~ dy = \int_0^h \pad u x ~ dy + u(h) h_x x ~ .
\eea
Hence
\begin{equation}
u(h)\frac{\partial h}{\partial  x} + \frac{\partial  h}{\partial  t} = -\left[ \pad{}{x} \int_0^h u ~ dy - u(h) h_x\right]  ~.
\end{equation}
Defining the fluid flux
\begin{equation}\label{flux}
Q = \int_{0}^{h(x)} u\, dy = \frac{\sigma}{3\mu} h_{xxx}h^3 + UH
\end{equation}
we obtain the well-known mass balance
\bea
\pad{h}{t} + \pad{Q}{x} = 0 ~ .
\eea

Then, taking the derivative with respect to $x$ of \eqref{flux} gives
\begin{equation}\label{dflux}
\frac{\partial Q}{\partial x} = \int_{0}^{h(x)} \frac{\partial u}{\partial x}\, dy + h_x u
\end{equation}
Finally, combining \eqref{comb} and \eqref{dflux} we obtain the general equation for the film hight
\begin{equation}
\frac{\partial h}{\partial t} + \frac{\partial Q}{\partial x} = 0\,.
\end{equation}

\section*{Appendix B: Flat or curved surface?}
Obviously the geometry of the problem involves radial symmetry, indicating we should analyse the governing equations in a radial system. However when the liquid layer is thin it is often possible to work in a Cartesian system. To see whether this is justified in the present situation we consider the stress on the paste surface. The stress is proportional to the curvature and in polar co-ordinates is given by
$$
\kappa = \frac{(R+h)_{xx}}{(1+(R+h)_x^2)^{3/2}} + \frac{1}{R(1+(R+h)_x^2)^{1/2}} ~ .
$$
The first term on the right corresponds to the height variation along the axis, the second term is the circular stress. This is the stress that will pinch off a jet of water.The radius $R$ is the constant radius of a wire, while $h$ is the variable paste thickness, noting that $R$ is constant we may write
$$
\kappa = \frac{h_{xx}}{(1+h_x^2)^{3/2}} + \frac{1}{R(1+h_x^2)^{1/2}} \sim \frac{H}{L^2} + \frac{1}{R} \sim 10^5 + 10^3 
$$
The first term represents the stress that would occur on a film of height $h$ on a flat surface, the second the circular stress. Given that the circular stress is 100 times smaller than the flat surface stress we may then ignore radial curvature (leading to errors of the order 1\%).
Consequently in this report we will deal only with the flat substrate equations.

\bibliographystyle{plain}
\bibliography{biblio_frenos.bib}

\end{document}